\documentclass[11pt]{article}

\usepackage[english]{babel}
\usepackage[defs]{ams}
\usepackage{hyperref}
\usepackage{amsmath}
\usepackage{amssymb,amsthm,amscd}
\usepackage[latin1]{inputenc}
\usepackage[textwidth=16cm,textheight=22cm]{geometry}
\usepackage{mathrsfs}

\newcommand{\diag}{\operatorname{diag}}

\newcommand{\n}{\mathcal{N}}
\newcommand{\m}{\mathcal{M}}

\newcommand{\g}{\mathfrak{g}}

\newcommand{\I}{\mathbb{I}}
\renewcommand{\d}{\mathrm{d}}

\newcommand{\bn}{{\bar n}}
\newcommand{\bm}{{\bar m}}

\newcommand{\Exp}[1]{\operatorname{e}^{#1}}

\renewcommand{\L}{\mathscr L}

\newcommand{\Ss}{\mathscr S}
\newtheorem{pro}{Proposition}

\newtheorem{theorem}{Theorem}

\begin{document}

\title{The multicomponent 2D Toda hierarchy: generalized matrix orthogonal polynomials, multiple orthogonal polynomials and Riemann--Hilbert problems}

\author{Carlos \'{A}lvarez-Fern\'{a}ndez$^\dag$, Ulises Fidalgo$^\ddag$ and Manuel Ma\~{n}as$^\dag$\\ \\
$^\dag$ Departamento de F\'{\i}sica Te\'{o}rica II, Universidad Complutense, \\
28040-Madrid, Spain\\
$^\ddag$ Departamento de Matem\'{a}tica Aplicada, Universidad Carlos III\\ 28911-Madrid, Spain }
\date{manuel.manas@fis.ucm.es}
\maketitle
\begin{abstract}
We consider the relation of the multi-component 2D Toda hierarchy with matrix orthogonal and biorthogonal polynomials. The multi-graded H\"{a}nkel reduction of this hierarchy is considered and the corresponding generalized matrix orthogonal polynomials are studied. In particular for these polynomials we consider the recursion relations, and for rank one weights its relation with multiple orthogonal polynomials of mixed type with a type II normalization and the corresponding link with a Riemann--Hilbert problem.
\end{abstract}

\section{Introduction}

Recently, multiple orthogonal polynomials, the related Riemann--Hilbert problems and its application to different areas, for example Brownian motions have deserved much attention \cite{daems-kuijlaars}. The field of matrix orthogonal polynomials has been also a growing area of research with some similarities with the scalar case but much more richness, see \cite{duran}.
The relation of multiple orthogonal polynomials with multicomponent KP hierarchies has been noticed in \cite{adler-vanmoerbeke2} and the string equation formalism of integrable systems
has been applied in \cite{luis}. Ten years ago   Adler and van Moerbeke \cite{adler-vanmoerbeke} --in the context of the so called discrete KP hierarchy-- introduced  what they named as generalized orthogonal polynomials, and what they claimed to be the corresponding Riemann--Hilbert problem, while later on they studied the related  Darboux transformations \cite{adler-vanmoerbeke3}. Recently, and for the Toeplitz case  Cafasso \cite{cafasso} extended this work in order to consider block matrices within the non-Abelian Ablotiwz--Ladick latttice.

Following Ueno and Takasaki \cite{ueno-takasaki} and the seminal paper of Mulase \cite{mulase} in \cite{manas-martinez-alvarez} we gave a description of the infinite multicomponent 2D Toda lattice hierarchy in terms of a Gaussian factorization (also known as Borel factorization) in an infinite dimensional Lie group, while later on \cite{manas2} we analyzed the dispersionless limit of the hierarchy with the aid of the factorization problem.  (See \cite{felipe-onjay} for a discussion on different cases where this factorization makes sense.) Following \cite{adler-vanmoerbeke} we could argue as follows: i) in the one hand the multicomponent Toda hierarchy may be viewed as an $LU$ factorization of certain deformed infinite-dimensional matrix and ii) on the other hand, the same matrix could be thought as a moment matrix and the corresponding $LU$ factorization should give us the corresponding generalized matrix orthogonal polynomials. In this manner,  we would be able to build a bridge between multicomponent Toda hierarchy and matrix orthogonal polynomials. This  is the main idea developed in this paper. First, we connect matrix orthogonal and biorthogonal polynomials with the multi-component 2DToda lattice hierarchy, focusing in particular on the H\"{a}nkel reduction. Second, we generalize the band condition of \cite{adler-vanmoerbeke} to the multicomponent case and consider what we refer as multi-graded H\"{a}nkel. This leads to a multi-component extension of  generalized orthogonal polynomials which in some cases can be described in terms of multiple orthogonal polynomials of mixed type with a   type II normalization. This connection allows us to give an appropriate Riemann--Hilbert problem for this generalized orthogonal polynomials (notice that the one discussed in \cite{adler-vanmoerbeke} is not correct).

The layout of the paper is as follows. In this introduction we give a overview of the Gaussian factorization and the semi-infinite multi-component 2D Toda lattice hierarchy. Then, in \S 2 we discuss how matrix orthogonal polynomials and the multi-component 2D Toda lattice hierarchy are connected. Finally, in \S 3 we consider multi-graded H\"{a}nkel reduction, extended generalized orthogonal polynomials, mixed multiple orthogonal polynomials and corresponding Riemann--Hilbert problems.


\subsection{Gaussian factorization and the semi-infinite multi-component 2D Toda lattice hierarchy}

For the construction of a Lie group theoretical setting we  denote  by $\Lambda$
the shift operator for matrix valued sequences. The associative
algebra of linear operators on these sequences can be identified with
the associative algebra of semi-infinite matrices with entries taking values in $\C^{N \times N}$, the set of $N\times N$
complex matrices. 
With the usual commutator for linear operators
this algebra is also a Lie algebra denoted by $\g$ whose Lie group
$G$  is the group of invertible linear operators in $\g$. Let's take
$g \in G$ and consider the following Gaussian factorization problem
$g=S^{-1}\bar S$ where $S$ is a block lower triangular matrix, with $S_{ii}=\I_N$, being $\I_N\in\C^{N\times N}$ the identity matrix,   and $\bar S$
is an block upper triangular matrix. In \cite{felipe-onjay}  it is proven that the Borel decomposition holds if all the principal minors do not vanish.
Thus, the factorization
holds under ``small'' continuous deformations and we can consider the  factorization  $g(t)=S(t)^{-1}\bar S (t)$ where
$t$ stands for a set  of  complex variables. As was discussed in \cite{manas-martinez-alvarez} this factorization problem leads to an
integrable hierarchy of nonlinear PDE known as multicomponent 2D Toda lattice hierarchy. Let us discuss these issues in more depth.
Observe that the matrix associated with the shift operator is the block matrix
$(\I_N \delta_{i,i+1})$ and $\Lambda^t$ is the operator
associated with the transposed matrix $(\I_N
\delta_{i+1,i})$. If $E_{ab}$, $a,b=1,\dots,N$ is the canonical basis of $\C^{N\times N}$ and $t=(\{t_{ja}\},\{\bar t_{j a}\})$, $j=1,2,\dots$ and $a=1,\dots,N$, is a collection of complex parameters we introduce
\begin{equation*}
    W_{0}:=\sum_{a=1}^N E_{aa} \exp(\sum_{j=1}^{\infty} t_{ja}
    \Lambda^j),
\end{equation*}
\begin{equation*}
  \bar W_{0}:=\sum_{a=1}^N E_{aa} \exp(\sum_{j=1}^{\infty} \bar t_{j a}
    (\Lambda^t)^j),
\end{equation*}
and  consider Gaussian  factorization of $ g(t):=W_{0}(t)g\bar W_{0}(t)^{-1}$.

Following \cite{manas-martinez-alvarez} we define the Lax operators
\begin{align}\label{lax}
\begin{aligned}
  L&:=S\Lambda S^{-1}=\Lambda+u_0+u_1\Lambda^{t}+u_2(\Lambda^{t})^2+\cdots,\\
  C_a&:=SE_{aa}S^{-1}=E_{aa}+c_{a1}\Lambda^{t}+\cdots,\\
 \bar L&:= \bar S \Lambda^t\bar S^{-1}=\Exp{\phi}\Lambda^t+\bar u_0+\bar u_1\Lambda+\cdots,\\
 \bar C_a&:= \bar S E_{aa}\bar S^{-1}=\bar c_{a0}+\bar c_{a1}\Lambda+\cdots,&
\end{aligned}
\end{align}
where all the coefficients in the  $\Lambda$-expansions belong to $\C^{N\times N}$.
The multi-component 2D Toda hierarchy has the following Lax representation
\begin{align*}
  \frac{\partial L}{\partial t_{ja}}&=[(L^jC_a)_+,L],&  \frac{\partial C_b}{\partial t_{ja}}&=[(L^jC_a)_+,C_b],&\frac{\partial \bar L}{\partial t_{ja}}&=[(L^jC_a)_+,\bar L],&  \frac{\partial \bar C_b}{\partial t_{ja}}&=[(L^jC_a)_+,\bar C_b],\\
  \frac{\partial L}{\partial \bar t_{ja}}&=[(\bar L^j\bar C_a)_-,L],&  \frac{\partial C_b}{\partial \bar t_{ja}}&=[(\bar L^j\bar C_a)_-,C_b],&\frac{\partial \bar L}{\partial \bar t_{ja}}&=[(\bar L^j\bar C_a)_-,\bar L],&  \frac{\partial \bar C_b}{\partial \bar  t_{ja}}&=[(\bar L^j\bar C_a)_-,\bar C_b],
  \end{align*}
where the sub-indices $+$ and  $-$ denote the block upper triangular, strictly block lower triangular projections, respectively.

\section{Matrix orthogonal polynomials and the multi-component 2DToda lattice hierarchy}
 Following Adler and van Moerbeke \cite{adler-vanmoerbeke} we construct families of matrix orthogonal and bi-orthogonal polynomials associated with the 2D Toda lattice hierarchy.

In the first place, we define the following families of (time-dependent) matrix
polynomials
\begin{align*}
  p(z)\equiv\{p_i(z)\}_{i\geq 0}&:=S\chi(z), &\bar
p(z)\equiv\{\bar p_i(z)\}_{i\geq 0}:=(\bar S ^{-1} )^{\dagger}
 \chi(z),&
\end{align*}
where
$\chi(z):=(\I_N,z\I_N,z^2\I_N,\ldots)^t$ and the symbol $^\dagger$ denotes
Hermitian conjugation.Next we consider a matrix-valued bilinear pairing between
matrix polynomials. Given matrix polynomials $P(z)=\sum_{k=0}^i P_k
z^k$ and $Q(z)=\sum_{l=0}^j Q_l z^l$ (of degrees $i,j$, respectively) we have
\begin{equation*}
\langle P(z), Q(z) \rangle = \sum_{\substack{k=1,\dots,i\\l=1,\dots,j}} P_{k} \langle
z^k\I_N,z^l\I_N \rangle Q_{l}^{\dagger},
\end{equation*}
where $\langle z^k\I_N,z^l\I_N \rangle$ denotes the matrix for the bilinear
pairing in the canonical basis and for each $(k,l)$ is an $N \times
N$ complex matrix.

This pairing has the following properties
\begin{enumerate}
  \item Is linear in the first component:
\begin{align*}
\langle{c_1P_1(z)+c_2P_2(z),Q(z)}\rangle&=c_1\langle{P_1(z),Q(z)}\rangle+c_2\langle{P_2(z),Q(z)}\rangle,&\forall c_1,c_2&\in\C^{N\times N}
\end{align*}
 \item Is skew-linear in the second component:
\begin{align*}
    \langle{P(z),c_1Q_1(z)+c_2Q_2(z)}\rangle&=\langle{P(z),Q_1(z)}\rangle
c_1^{\dagger}+\langle{P(z),Q_2(z)}\rangle c_2 ^{\dagger},&\forall c_1,c_2&\in\C^{N\times N}
\end{align*}
\end{enumerate}

\begin{pro}\label{pro1}
\begin{enumerate}
\item If $\langle z^i\I_N,z^j\I_N\rangle=g(t)_{ij}$ where $g_{ij}$ is the
$\C^{N \times N}$ block in the position $(i,j)$, then the families $p(z)$
and $\bar p(z)$ are biorthogonal matrix polynomials for the linear
pairing, i.e.
\begin{align*}
  \langle p_i(z), \bar p_j(z) \rangle=\delta_{ij}\I_N.
\end{align*}
Moreover,
\begin{align}\label{orthogonality}
  \langle p_i(z), z^l\I_N\rangle&=0,& l&=0,\dots, i-1,&
  \langle z^l\I_N,\bar p_j(z) \rangle&=0,& l&=0,\dots, j-1.&
\end{align}
\item In addition, if the time-dependent initial condition $g(t)$ is Hermitian
 for all $t$ then $p(z)$ and $\bar p(z)$ are two families
of matrix orthogonal polynomials, moreover, the two families are
proportional.
\end{enumerate}

\end{pro}
\begin{proof}
\begin{enumerate}
\item With the previous definitions for $p(z)$ and $\bar p(z)$ we have:
\begin{align*}
    p_i(z)&=\sum_{k=0}^i S_{ik}z^k, &
   \bar p_j(z)&=\sum_{l=0} ^j (\bar S_{lj}^{-1})^{\dagger}z^l,
\end{align*}
where $S_{ik}$ and $\bar S_{lj}^{-1}$ are the blocks $(i,k)$ and
$(l,j)$ for $S$ and $\bar S^{-1}$, respectively. Hence
\begin{align*}
\langle p_i(z), \bar p_j(z) \rangle&=\sum_{k,l=0}^{i,j} S_{ik}
\langle z^k\I_N,z^l\I_N \rangle \bar S_{lj}^{-1}=\sum_{k,l \geq 0} S_{ik}
\langle
z^k\I_N,z^l\I_N \rangle \bar S_{lj}^{-1}\\
&=(S g(t) \bar S^{-1})_{ij}=(S S^{-1} \bar S \bar S
^{-1})_{ij}=\delta_{ij}\I_N,
\end{align*}
as desired.
Finally, \eqref{orthogonality} is proven by induction. First we have that $\langle p_i(z),\bar p_0(z)\rangle=0$, but $p_0(z)=(\bar S_{00}^{-1})^{\dagger}$ is invertible and therefore
we conclude $\langle p_i(z),\I_N\rangle=0$. Now,  $\langle p_i(z),\bar p_1(z)\rangle=0$, but $\bar p_1(z)=(\bar S_{11}^{-1})^{\dagger}z+(\bar S_{10}^{-1})^{\dagger}$, and using the skew-linearity, the previous result  and the fact that $(\bar S_{11}^{-1})^{\dagger}$ is invertible we deduce that  $\langle p_i(z),z\I_N\rangle=0$, and so forth an so on.

\item  Let's study the conditions under which $p(z)$ is a family of matrix
orthogonal polynomials. If we take two polynomials in the family,
such as $p_i(z)=\sum_{k=1}^i S_{ik}z^k$ and $p_j(z)=\sum_{l=1}^j
S_{jl}z^l$ we have
\begin{align*}
    \langle p_i(z),p_j(z) \rangle &=\sum_{k,l=1}^{i,j} S_{ik} \langle z^kI,z^lI
    \rangle (S_{jl})^{\dagger}=\sum_{k,l=0}^{\infty} S_{ik} \langle z^k\I_N,z^l\I_N
    \rangle (S^{\dagger})_{lj}\\
    &=(Sg(t)S^{\dagger})_{ij}=(\bar S S^{\dagger})_{ij}.
\end{align*}
Observe that $\bar S S^{\dagger}$ is clearly block upper-diagonal with its Hermitian conjugate given by
\begin{align*}
(\bar S S^{\dagger})^{\dagger}=S \bar S ^{\dagger}=S g(t)
(g(t)^{-1})^{\dagger}\bar S ^{\dagger}=\bar S (\bar S
g(t)^{-1})^{\dagger}=\bar S S^{\dagger}.
\end{align*}
Therefore,  $\bar S S^{\dagger}$ is Hermitian and block
upper-diagonal, which implies that $\bar S S^{\dagger}$ is a block
diagonal matrix and the blocks in the diagonal are $\C^{N \times N}$
Hermitian matrices.

We conclude that $\langle p_i(z),p_j(z)
\rangle=\delta_{ij}(h_i)^{-1}$, where $h_i$ is a Hermitian matrix.
Notice also that as a consequence  $\bar p(z)=h p(z)$ where
$h=\text{diag}(h_1,h_2,\ldots)$.
\end{enumerate}
\end{proof}

\subsection{The H\"{a}nkel case}
We choose $g$ to be a block H\"{a}nkel matrix so that $\Lambda g=g \Lambda^{t}$ or
\begin{align*}
    (\Lambda g)_{ij}=g_{i+1,j}=g_{i,j+1}=(g \Lambda^t)_{ij},
\end{align*}
In this case we have for the blocks of the moment matrix $g$
\begin{align*}
  g_{ij}=\gamma^{(i+j)},
\end{align*}
for some matrices $\gamma^{(j)}\in\C^{N\times N}$.
From
\begin{align*}
\Lambda g(t)&=\Lambda W_{0}g\bar W_{0}^{-1}=W_{0}\Lambda
g\bar W_{0}^{-1}\\
&=W_{0}g \Lambda^t \bar W_{0}^{-1}=W_{0} g
\bar W_{0}^{-1} \Lambda^t=g(t)\Lambda^t,
\end{align*}
we easily deduce that  $g(t)$ is  block H\"{a}nkel if $g$ is.
 We have
 \begin{pro}
 Assume that  $g$ is block H\"{a}nkel then
with $\gamma^{(j)}$  a block moment matrix;  i.e. $\gamma^{(j)}=\int_{\R}x^j\rho(x)\d x$ then the pairing can be viewed as a scalar product in the real line whose matrix moment
is $g$ ; i.e.,
\begin{align*}
    \langle P(x),Q(x) \rangle=\int_{\R} P(x) \rho(x) Q(x)^{\dagger} \d x.
\end{align*}
 \end{pro}
\begin{proof}
On one hand we have $\langle P(x), Q(x) \rangle=\sum_{ij} P_i
\gamma^{(i+j)} Q_j^{\dagger}$. Using the previous definition $\int_\R
x^{j+k}\rho(x)\d x= \langle x^j\I_N,x^k\I_N \rangle=\gamma^{(j+k)}$ (the
H\"{a}nkel symmetry ensures that there is only dependence in $j+k$) we
have:
\begin{align*}
\langle P(x), Q(x) \rangle&=\sum_{ij} P_i \gamma^{(i+j)}
Q_j^{\dagger}=\sum_{ij} P_i \int_\R x^{j+k}\rho(x)\d x
Q_j^{\dagger}\\&=\int_{\R} P(x) \rho(x) Q(x)^{\dagger} \d x.
\end{align*}

\end{proof}

In general arbitrary continuous deformations do not preserve the
Hermitian character of  $g$. If we look for families of matrix orthogonal
polynomials on the real line we should make restricted deformations. Let's make this
point clear. In the following $z^*$ denotes the complex conjugate of $z\in\C$.

\begin{pro}
   If the matrix $g$ is block H\"{a}nkel and the matrices $\gamma^{(j)}$
are Hermitian then
\begin{enumerate}
\item The families $p(z)$ and $\bar p(z)$ are proportional
and the two of them are matrix orthogonal polynomials in the real
line.
\item Moreover, if the continuous deformation parameters satisfy one of the two
following conditions
\begin{enumerate}
  \item $t_{ja},\bar t_{j a}\in\R$ and satisfy
  $t_{ja}=t_{j},\bar t_{j a}=\bar t_{j}$, $a=1,\dots,N$.
  \item $t_{ja},\bar t_{j  a}$ satisfy $t_{ja}+{\bar t_{j
  a}}^{\ast}=0$.
\end{enumerate}
the result holds for the time dependent moment matrix.
\end{enumerate}
\end{pro}
\begin{proof}
\begin{enumerate}
\item If the matrix $g$ is block H\"{a}nkel and the blocks are Hermitian the
matrix $g$ is itself Hermitian. Given $i,j$, pair of indices for an
element of $g$, there exist four integer indices $(k,l)$,$(m,n)$ with
$k,l \geq 0$ y $ m,n =1,\dots, N$ that satisfy
$a_{ij}=(A_{kl})_{mn}=(A_{kl})_{nm}^{\ast}=(A_{lk})_{nm}^{\ast}=a_{ji}^{\ast}$
(we use $A\in\C^{\N \times N}$ for a block of $g$) as a consequence
$g=g^{\dagger}$ and by Proposition \ref{pro1} the first part of the result holds.
\item Let us  compute the time evolution
for $g$.
We have
\begin{equation*}
W_{0}=\sum_{a=1}^N E_{aa} \exp(\sum_{j=1}^{\infty} t_{ja}
    \Lambda^j)=\sum_{a=1}^N E_{aa} \sum_{j=0}^{\infty} s_j^{(a)}
    \Lambda^j,
\end{equation*}
where $s_j^{(a)}$ is the $j$-th Schur\footnote{We remind the reader that the Schur polynomials are given by $\exp({\sum_{j=1}^{\infty}t_{ja}\Lambda^j})=\sum_{j=0}^{\infty}s_j^{(a)}\Lambda^j$ so $s_0^{(a)}=1$ and $s_j^{(a)}=\sum_{p=1}^j\sum_{j_1+\cdots+j_p=j} t_{j_1a}\cdots t_{j_p a}
=t_{ja}+\cdots$ where the missing term is a polynomial of utmost degree $j$ in the variables $t_{1a},t_{2a},\ldots,t_{j-1,a}$.}
polynomial for the component $a$. Consequently and taking $s_j:=\sum_{a=1}^NE_{aa}s_j^{(a)}$
\begin{equation*}
    W_{0}=\sum_{j=0}^{\infty} \sum_{k=1}^N E_{aa}  s_j^{(a)}
    \Lambda^j=\sum_{j=0}^{\infty} s_j
    \Lambda^j,
\end{equation*}
so it is straight forward to see the block structure of $W_0$, whose blocks are given by $(W_0)_{ij}=s_{j-i}$ if $j-i \geq 0$ and $0_N$ otherwise.

A similar argument for $\bar W_{0}^{-1}$ leads to $(\bar W_0^{-1})_{ij}=\bar s_{i-j}$ if $i-j \geq 0$ and $0_N$ otherwise (here $\bar s_j:=\sum_{a=1}^NE_{aa}\bar s_j^{(a)}$ and $\bar s_j^{(a)}$ is the $j$-th Schur polynomial for the component $a$ but now in the variables $-\bar t_{ja}$).

We are now ready to compute
\begin{equation*}
    g(t)_{ij}=\sum_{k,l\geq 0}(W_0)_{ik}g_{kl}(\bar W_0^{-1})_{lj}=\sum_{k \geq i, l\geq j} s_{k-i}\gamma^{(k+l)} \bar s _{l-j}=
    \sum_{k,l \geq 0} s_{k}\gamma^{(i+j+k+l)} \bar s _{l}.
\end{equation*}

Then
\begin{enumerate}
\item  If the first condition holds then all matrices $s_j$ and $\bar s_j$ are real and scalar so
\begin{align*}
    g(t)_{ij}^{\dagger}&=\sum_{k,l \geq 0} \bar s_l^{\dagger} \gamma^{(i+j+k+l)}
    s_k^{\dagger}=\sum_{k,l \geq 0} s_k^{\dagger} \gamma^{(i+j+k+l)}
    \bar s_l^{\dagger}=\sum_{k,l \geq 0} s_k \gamma^{(i+j+k+l)}
    \bar s_l\\&=g(t)_{ij}.
\end{align*}
\item If the second condition holds $s_j=\bar s_j^{\dagger}$ for all $j$ so
\begin{align*}
    g(t)_{ij}^{\dagger}&=\sum_{k,l \geq 0} \bar s_l^{\dagger} \gamma^{(i+j+k+l)}
    s_k^{\dagger}=\sum_{l,k \geq 0} \bar s_k^{\dagger} \gamma^{(i+j+l+k)}
    s_l^{\dagger}=\sum_{k,l \geq 0} s_k \gamma^{(i+j+k+l)}
    \bar s_l\\&=g(t)_{ij}.
\end{align*}
\end{enumerate}
Under any of the two conditions $g(t)$ is block-H\"{a}nkel with Hermitian blocks, so is itself Hermitian. That proves the second part of the proposition.
\end{enumerate}
\end{proof}

We now discuss the recursion formulae. Using the H\"{a}nkel condition $\Lambda g=g\Lambda^t$ and taking the
usual definition for the Lax operator we conclude $L=\bar L$ and hence
 we have $L=\Lambda+u(n)+v(n)\Lambda^t$.
\begin{pro}
 The polynomials $p(z),\bar p(z)$ satisfy a three term recurrence law
 given by $p_{n+1}(z)=zp_n(z)-u(n)p_n(z)-v(n)p_{n-1}(z)$
\end{pro}
\begin{proof}
  From the definition of $L$ we have
$Lp(z)=LS\chi(z)=zp(z)=(\Lambda+u(n)+v(n)\Lambda^t)p(z)$.
If we take the sequence terms $\{p_n(z)\}_n$ we conclude
$zp_n(z)=p_{n+1}(z)+u(n)p_n(z)+v(n)p_{n-1}$.
For the polynomials $\bar p(z)$ we have $\bar L^{\dagger} \bar
p(z)=z\bar p(z)=(\Lambda
v^{\dagger}(n)+u^{\dagger}(n)+\Lambda^t)\bar p(z)$, and using the
sequence $\{\bar p_n(z)\}_n$ we obtain another recurrence law.
\end{proof}

\section{Multi-graded H\"{a}nkel reduction, generalized orthogonality, multiple orthogonal polynomials and Riemann--Hilbert problems}
In this section we will study generalized H\"{a}nkel type conditions. Given a multi-index $\bn=(n_1,\dots,n_N)$ with $n_a$ non-negative integers we define for $A\in\g$ the power $A^\bn=\sum_{a=1}^N A^{n_a}E_{aa}$. For two multi-indices $\bn$ and $\bm$  a matrix $g$ is said to be a $(\bn,\bm)$ multi-graded H\"{a}nkel  if
\begin{align}\label{mhankel}
    \Lambda^\bn g= g (\Lambda^t)^\bm.
\end{align}
If as before $g_{ij}\in\C^{N\times N}$ denotes a block in $g$ then we can write $g_{ij}=(g_{ij,ab})_{1\leq a,b\leq N}$ and the multi-graded H\"{a}nkel condition reads
$g_{i+n_a\,j,ab}=g_{i\, j+m_b,ab}$. An ample family of multi-graded H\"{a}nkel matrices can be constructed in terms of weights $\rho_{j,ab}$ as the moments
\begin{align}\label{mghankel}
  g_{ij,ab}=\int_\R x^i\rho_{j,ab}(x)\d x,
\end{align}
where the weights satisfy a generalized periodicity condition of the form
\begin{align}\label{periodicity}
  \rho_{j+m_b,ab}(x)=x^{n_a}\rho_{j,ab}(x).
\end{align}
Thus, given the weights $\rho_{0,ab},\dots,\rho_{m_b-1,ab}$,
all the others are fixed by \eqref{periodicity}. From now on, we concentrate only in these cases of multigraded H\"{a}nkel matrices. We  notice that for the 1-component case and for $n_1=m_1$ case these moments matrices were studied in \cite{adler-vanmoerbeke}-\cite{adler-vanmoerbeke3}.

\begin{pro}
  For multi-graded H\"{a}nkel matrices the matrix polynomials $p_i$ satisfy the following generalized orthogonality conditions
\begin{align}\label{generalized orthogonality}
 \int_{\R}p_i(x)\rho_j(x)\d x&=0,& j&=0,\dots,i-1,&    \rho_j&:=(\rho_{j,ab})\in\C^{N\times N}.
\end{align}
\end{pro}
\begin{proof}
  From $Sg(t)\bar S^{-1}=1_G$ we get $\sum_{\substack{j=1,\dots,i\\ b=1,\dots,N}}p_{ij,ab}g_{jl,bc}=0$ for $a,c=1,\dots,N$ and $l=0,\dots,i-1$. Now recalling \eqref{mghankel} we get the result.
\end{proof}

 Using the Euclidean division  $ i=\theta_cm_c+\sigma_c$,  with $\theta_c\geq 0$, $0\leq \sigma_c< m_c$ we get a better insight of the orthogonality relations \eqref{generalized orthogonality}, for $a,c=1,\dots,N$,
\begin{align}
  \begin{aligned}
    \sum_{b=1}^N \int_\R p_{i,ab}(x)\rho_{j,bc}(x)(x^{n_b})^l\d x&=0, &j=0,\dots,m_c-1,&&l=0,\dots, \theta_c-1,\\
     \sum_{b=1}^N \int_\R p_{i,ab}(x)\rho_{j,bc}(x)(x^{n_b})^{\theta_c}\d x&=0, &j=0,\dots,\sigma_c-1.
  \end{aligned}  \label{gop-components}
\end{align}

\subsection{Evolution}

From \eqref{mhankel} we conclude that $g(t)$ is of $(\bar n,\bar m)$ multi-graded H\"{a}nkel type if $g$ is. In general, the evolution of $g$  is given in terms of Schur polynomials by
\begin{align*}
g_{ij}(t)=\sum_{k,l\geq 0}s_kg_{i+k,j+l}\bar s_l
\end{align*}
which recalling \eqref{mghankel} leads to the following evolution of the weights
\begin{align}\label{evol.mg}
  \rho_j(t)=\sum_{k,l\geq 0}x^ks_k\rho_{j+l}\bar s_l=\exp(t(x))\sum_{l\geq 0}\rho_{j+l}\bar s_l,
\end{align}
where $t(x)=\sum_{a=1}^Nt_a(x)E_{aa}$ and $t_a(x):=\sum_{j\geq 1} t_{ja}x^j $.
It can be  easily checked that  $\rho_j(t)$ satisfies the periodicity condition \eqref{periodicity} if $\rho_j$ does. From \eqref{evol.mg} we infer that
\begin{align}
  \begin{pmatrix}
    \rho_{0,ab}(t)\\\vdots\\
    \rho_{m_b-1,ab}(t)
  \end{pmatrix}=\exp(t_a(x))
\begin{pmatrix}
  \Ss_{0,ab}&\Ss_{1,ab}&\Ss_{2,ab}&\cdots&\Ss_{m_b-1,ab}\\
   x^{n_a}\Ss_{m_b-1,ab}&\Ss_{0,ab}&\Ss_{1,ab}&\cdots&\Ss_{m_b-2,ab}\\
     x^{n_a}\Ss_{m_b-2,ab}& x^{n_a}\Ss_{m_b-1,ab}&\Ss_{0,ab}&\cdots&\Ss_{m_b-2,ab}\\
    \vdots&\vdots&\vdots&\ddots &\vdots     \\
     x^{n_a} \Ss_{1,ab}&x^{n_a}\Ss_{2,ab}&x^{n_a}\Ss_{3,ab}&\cdots&\Ss_{0,ab}
\end{pmatrix}
 \begin{pmatrix}
    \rho_{0,ab}\\\vdots\\
    \rho_{m_b-1,ab}
  \end{pmatrix}
\end{align}
where
\begin{align*}
  \Ss_{i,ab}&=\sum_{j\geq 0}\bar s_{i+m_bj} ^{(b)}(x^{n_a})^j=\frac{1}{m_b x^{in_a/m_b}}\sum_{k=0}^{m_b-1}\varepsilon_b^{ik}\exp(-\bar t_b(\varepsilon_b^kx^{n_a/m_b})),& \varepsilon_b^{m_b}=1.
\end{align*}
If we denote
\begin{align*}
  \bar t_a^{(l)}(x)&=\sum_{j\geq 0}\bar t_{jm_b+l,a}x^{jm_b+l},&l&=0,1,\dots,m_b-1,
\end{align*}
we have
\begin{align*}
  \bar t_a(\varepsilon_b^k x)&=\bar t_a^{(0)}(x)+\bar t_a^{[k]}( x),& \bar t_a^{[k]}( x)&=\sum_{\substack{j\geq 0\\l=1,\dots, m_b-1}}\bar t_{jm_b+l}\varepsilon_b^{kl}x^{jm_b+l},
\end{align*}
and therefore
\begin{align}\label{ecuacion postiva}
  \Ss_{i,ab}&=\frac{1}{m_b x^{in_a/m_b}}\exp(-\bar t_b^{(0)}(x^{n_a/m_b}))\sum_{k=0}^{m_b-1}\varepsilon_b^{ik}\exp(-\bar t_b^{[k]}(x^{n_a/m_b})).
\end{align}
Finally, we deduce
\begin{align}
  \label{rho.evol}
  \rho_{j,ab}(t)=\exp(t_a(x)-\bar t_b^{(0)}(x^{n_a/m_b}))\sum_{k=0}^{m_b-1}\hat\rho^{(k)}_{j,ab}\exp{(-\bar t_b^{[k]}(x^{n_a/m_b}))},
\end{align}
where we have used the discrete Fourier transform of the weights
\begin{align*}
\hat\rho^{(k)}_{j,ab}:=\frac{1}{m_b}\sum_{i=1}^{m_b-1}\varepsilon_b^{ik}x^{-in_a/m_b}\rho_{j+i,ab}.
\end{align*}
\subsection{Recursion relations and symmetries}

In terms of Lax operators the multi-graded H\"{a}nkel reduction reads \cite{manas-martinez-alvarez}
\begin{align}\label{recurrence1}
\L:= \sum_{a=1}^N L^{n_a}C_a= \sum_{b=1}^N \bar L^{m_b}\bar C_b.
\end{align}
Within this subsection we assume that
\begin{align*}
 & n_1\geq\dots\geq  n_N\geq 1, &   &m_1\geq \dots\geq  m_N\geq 1,
\end{align*}
and suppose that $n_1=\dots=n_r$ and $n_r>n_{r+1}$.
Given \eqref{lax} from  \eqref{recurrence1} we deduce that 
\begin{align*}
  \L=(E_{11}+\dots+E_{rr})\Lambda^{n_1}+\L_{n_1-1} \Lambda^{n_1-1}+\cdots+\L_0+\L_{-1}\Lambda^t+\L_{-2}(\Lambda^t)^2+\cdots,
\end{align*}
while we also have $\L = \sum_{b=1}^N \bar L^{m_b}\bar C_b$ and therefore
\begin{align*}
  \L=\L_{-m_1}(\Lambda^{t})^{m_1}+\L_{-m_1+1} (\Lambda^t)^{m_1-1}+\cdots+\L_0+\L_{1}\Lambda+\L_{2}\Lambda^2+\cdots.
\end{align*}
We conclude the block band  structure
\begin{align}
  \label{Lform}
  \L=(E_{11}+\dots+E_{rr})\Lambda^{n_1}+\L_{n_1-1} \Lambda^{n_1-1}+\dots+\L_{-m_1+1} (\Lambda^t)^{m_1-1}+\L_{-m_1}(\Lambda^{t})^{m_1}.
\end{align}

\begin{pro}
  The polynomials $p_i(z)$ are subject to
  \begin{align*}
    (E_{11}+\dots+E_{rr})p_{i+n_1}(z)+\cdots
    +\L_{-m_1}p_{i-m_1}(z)=p_i(z)\big(\sum_{a=1}^n z^{n_a}E_{aa}\big)
  \end{align*}
\end{pro}
\begin{proof}
  We only have to show that $\L p=p(\sum_{a=1}^n z^{n_a}E_{aa})$. But this follows from $\L p=S(\sum_{a=1}^N\Lambda^{n_a}E_{aa})S^{-1}S\chi=
  S\chi(\sum_{a=1}^Nz^{n_a}E_{aa})=p(z)(\sum_{a=1}^n z^{n_a}E_{aa})$.
\end{proof}

Similarly, from $\L^\dagger \bar p(z)=\bar p(z)(\sum_{b=1}^N z^{m_b}E_{bb})$ a recursion relation follows for the $\bar p_i$.

Finally we notice that in this case the following symmetry conditions   hold \cite{manas-martinez-alvarez}
\begin{align*}
& \Big(\sum_{a=1}^N\frac{\partial }{\partial t_{i n_a\, a}}+\sum_{a=1}^N\frac{\partial }{\partial \bar t_{i m_a\, a}}\Big)L=  \Big(\sum_{a=1}^N\frac{\partial }{\partial t_{i n_a\, a}}+\sum_{a=1}^N\frac{\partial }{\partial \bar t_{i m_a\, a}}\Big)\bar L=0,\\
   &\Big(\sum_{a=1}^N\frac{\partial }{\partial t_{i n_a\, a}}+\sum_{a=1}^N\frac{\partial }{\partial \bar t_{i m_a\, a}}\Big)C_{l}=  \Big(\sum_{a=1}^N\frac{\partial }{\partial t_{i n_a\, a}}+\sum_{a=1}^N\frac{\partial }{\partial \bar t_{i m_a\, a}}\Big)\bar C_{b}=0,
 \end{align*}
for $i\geq 1$ and $b=1,\dots, N$.
\subsection{Relation with multiple orthogonal polynomials}

We will see that when  the weights $\rho_j$ are particular rank one matrices there is a nice correspondence with multiple orthogonal polynomial of mixed type with a normalization of type II.
Following  \cite{daems-kuijlaars} we  take two sets of non
negative multi-indices $\bar \nu=(\nu_1,\ldots,\nu_p)$, $\bar
\mu=(\mu_1,\ldots,\mu_q)$ and write $|\nu|=\sum_{J=1}^p \nu_J$ and
$|\mu|=\sum_{K=1}^q \mu_K$. We also take weights $\{w_{1J}\}_{J=1}^p$ and $\{w_{2K}\}_{K=1}^q$ which are assumed to be non-negative functions on the real line. For a fixed pair $\bar \nu, \bar \mu$ we say that
$\{A_{\bar \nu,\bar\mu, J}\}_{J=1,\ldots,p}$ is a set of multiple
orthogonal polynomials of mixed type if  $\deg A_{\bar \nu,\bar \mu, J}
\leq \nu_J-1$ and the following orthogonality relations
\begin{align}\label{mop}
    \int_{\R}\sum_{J=1}^p A_{\bar \nu, \bar \mu,
    J}(x)w_{1J}(x)w_{2K}(x)x^{\alpha}\d x&=0,& \alpha&=0,\ldots,\mu_K-1,& K=1,\dots,q,
\end{align}
are satisfied. Alternatively defining the following linear forms $Q_{\bar \nu, \bar
\mu}(x):=\sum_{J=1}^n A_{\bar \nu, \bar\mu,
    J}(x)w_{1J}(x)$ the orthogonality relations can be written in
    the following way
    \begin{align*}
        \int_{\R}Q_{\bar \nu, \bar
\mu}(x)w_{2K}(x)x^{\alpha} \d x &=0,& \alpha&=0,\ldots,\mu_K-1,& K=1,\dots,q.
    \end{align*}
    For multiple orthogonal polynomials as described above we will take $|\nu|=|\mu|$ and assume that the following conditions hold
\begin{enumerate}
  \item For each $K=1,\ldots,p$ the orthogonality relations for the
  multi-indices $\bar \nu+e_K,\bar \mu$ have a unique solution with $A_{\bar \nu,\bar \mu, K}$ monic and $\deg A_{\bar \nu,\bar \mu, K}
= \nu_K-1$, with $\deg A_{\bar \nu,\bar \mu, J} <\nu_J-1$ if $J\neq K$. We
will call it type II normalization to the $K$-th component and write
that normalized solution as $\{A_{\bar \nu, \bar \mu,
J}^{(\text{II},K)}\}_{J=1,\ldots,p}$.
  \item For each $K=1,\ldots,q$ the orthogonality relations for the
  multi-indices $\bar \nu,\bar \mu-e_K$ have a unique solution with the
  following normalization: $\deg A_{\bar \nu,\bar \mu, J}
= \nu_J-1$ and $\int_{\R}Q_{\bar \nu, \bar \mu}(x)w_{2K}x^{\mu_K} \d x =1$.
We will call it normalization of type I to the $K$-th component and
write that normalized solution as $\{A_{\bar \nu, \bar \mu,
J}^{(\text{I},K)}\}_{J=1,\ldots,p}$.
\end{enumerate}
    In order to connect \eqref{gop-components} with multiple orthogonal polynomials we consider the Euclidean division  $i=q_an_a+r_a$,with $q _a\geq 0$ and $0\leq r_a <n_a$,  and write
\begin{equation}
\label{defPi}
    p_{i,ab}(z)=\sum_{j=1}^{n_b} z^{j-1}\Pi_{ij,ab}(z^{n_b}),
\end{equation}
where $\Pi_{ij,ab}(z^{n_b})$ are polynomials in $z^{n_b}$ such that
\begin{align*}
  \deg \Pi_{ij,ab}\leq
  \begin{cases}
    q_b, & j\leq r_b \text{ or $j=r_b+1$ and $b=a$},\\
    q_b-1, & j>r_b+1 \text{ or $j= r_b+1$ and $b\neq a$.}
  \end{cases}
\end{align*}
Notice that the monic character of $p_i$ gives the normalization of  $\Pi_{i\;r_a+1,aa}$ which happens to be a monic polynomial with  $\deg
\Pi_{i\;r_a+1,aa}=q_a$.

The inversion formula for \eqref{defPi} can be deduced  as follows.
If we denote by $\epsilon_b:=\exp(2\pi\text{i}/n_b)$, a primitive $n_b$-th root of the unity, and evaluate at $\epsilon_b^k z$, $k=0,\ldots,n_b-1$ we get the following system of equations
\begin{align*}
    p_{i,ab}(\epsilon_b^k z)=\sum_{j=1}^{n_b} (\epsilon_b^kz)^{j-1}\Pi_{ij,ab}(z^{n_b}) \quad k=0,\dots,n_b-1,
\end{align*}
that we solve in order to obtain the polynomials $\Pi_{ik,ab}$ in terms of a discrete Fourier transform of the polynomial $p_i$ through the formula
\begin{align*}
  \Pi_{i\;j+1,ab}(z^{n_b})=\frac{1}{n_bz^j}\sum_{k=0}^{n_b-1} \epsilon_b^{-jk}p_{i,ab}(\epsilon_b^{k}z).
\end{align*}

Then, \eqref{gop-components} can be written as
 \begin{align}\label{bomba}
 \begin{aligned}
  \int_\R  \sum_{b=1}^N \sum_{j=1}^{n_b}\Pi_{ij,ab}(x^{n_b})x^{j-1}\rho_{k,bc}(x)(x^{n_b})^l\d x&=0, &k=0,\dots,m_c-1,&&l=0,\dots, \theta_c-1,\\
    \int_\R  \sum_{b=1}^N \sum_{j=1}^{n_b}\Pi_{ij,ab}(x^{n_b})x^{j-1}\rho_{k,bc}(x)(x^{n_b})^{\theta_c}\d x&=0, &k=0,\dots,\sigma_c-1.
 \end{aligned}
   \end{align}
 These equations strongly suggest to perform a change of variables in each integrand of the type $y=x^{n_b}$ . For that aim, it is relevant that when $n_b$ is an even number then $\text{supp} (\rho_{j,bc})\subset \R^+$, otherwise the change of variables is ill defined. In fact, for these even cases, one easily see that the weights must be supported on the positive axis or uniqueness of the orthogonal polynomials is not ensured. Moreover, for $n_b$ odd it is also necessary to assume that the weight is supported  either only in the positive real numbers or only in the negative real line, this requirement comes from the positivity condition on the weights and the use of \eqref{v}. Hereon we will assume that the all weights are supported on the  positive real semiline. When the mentioned change of variable is performed in \eqref{bomba}  we get
  \begin{align}\label{mop-gop1}  \begin{aligned}
   \int_\R  \sum_{b=1}^N \sum_{j=1}^{n_b}\Pi_{ij,ab}(y)\tilde\rho_{jk,bc}(y)y^l\d y&=0, &k=0,\dots,m_c-1,&&l=0,\dots, \theta_c-1,\\
    \int_\R  \sum_{b=1}^N \sum_{j=1}^{n_b}\Pi_{ij,ab}(y)\tilde\rho_{jk,bc}(y)y^{\theta_c}\d y&=0, &k=0,\dots,\sigma_c-1,
  \end{aligned}
  \end{align}
  with
    \begin{align*}
    \tilde\rho_{jk,bc}(y)&=\frac{1}{n_b}y^{\frac{j}{n_b}-1} \rho_{k,bc}\big(y^{\frac{1}{n_b}}\big).
  \end{align*}
 Now, if the matrix weights $\rho_k$ are rank-1 matrices of the following particular form
  \begin{align*}
    \rho_{k,bc}(x)=v_{1,b}(x)w_{2,kc}(x^{n_b}),
  \end{align*}
  we get
\begin{align}\label{mop-gop}  \begin{aligned}
   \int_\R  \sum_{b=1}^N \sum_{j=1}^{n_b}\Pi_{ij,ab}(y)w_{1,jb}(y)w_{2,kc}(y)y^l\d y&=0, &k=0,\dots,m_c-1,&&l=0,\dots, \theta_c-1,\\
    \int_\R  \sum_{b=1}^N \sum_{j=1}^{n_b}\Pi_{ij,ab}(y)w_{1,jb}(y)w_{2,kc}(y)y^{\theta_c}\d y&=0, &k=0,\dots,\sigma_c-1,
  \end{aligned}
  \end{align}
  where
  \begin{align}\label{v}
    w_{1,jb}(y)&=\frac{1}{n_b}y^{\frac{j}{n_b}-1} v_{1,b}\big(y^{\frac{1}{n_b}}\big).
  \end{align}

  We are ready to describe  the relation among multiple orthogonal polynomials of type II and  generalized matrix polynomials. First, given $(a,j)$ with $a=1,\dots,N$ and $j \in \N$ we make the definitions $\n(a,j):=n_1+\dots+n_{a-1}+j$ and $\m(a,j):=m_1+\dots+n_{a-1}+j+1$.
  \begin{pro}\label{identification}
  Relations \eqref{mop-gop} are particular cases of \eqref{mop} with: \begin{enumerate}
  \item $J=\n (b,j)$ for $b=1,\dots,N$ and $j=1,\dots,n_b$; and $K=\m (c,k)$ for $c=1,\dots,N$ and $k=0,\dots,m_c-1$.
  We have therefore the identification $J=n_1+\dots+n_{b-1}+j$ and $K=m_1+\dots+m_{c-1}+k+1$.
  \item $p=|\bn|=n_1+\dots+n_N$ and $q=|\bm|=m_1+\dots+m_N$.
  \item \begin{align*}
    \nu_{\n(b,j)}&=
\begin{cases}
    q_b+1, & j\leq r_b \text{ or $j=r_b+1$ and $b=a$},\\
    q_b, & j>r_b+1 \text{ or $j= r_b+1$ and $b\neq a$,}
  \end{cases}\\
\mu_{\m(c,k)}&=\begin{cases}
      \theta_c+1, & 0\leq k\leq \sigma_c-1,\\
      \theta_c,&\sigma_c\leq k\leq m_c-1.
    \end{cases}
  \end{align*}
\item $|\bar \nu|=\sum_J\nu_J=Ni+1$ and $|\bar \mu|=\sum_K\mu_K=Ni$.
\item $A_{\bar\nu,\bar\mu,J}=\Pi_{ij,ab}$ and $p_{i,ab}(z)=\sum_{j=1}^{n_b}z^{j-1}A_{\bar\nu,\bar\mu,\n(b,j)}(z^{n_b})$.
\item $\deg A_{\bar\nu,\bar\mu,J}=\nu_J$ if  $J=\n(a,r_a+1)$ so that $A_{\bar\nu,\bar\mu,J}=A_{\bar\nu,\bar\mu,J}^{(II,\n(a,r_a+1))}$, and we have a type $II$ normalization w.r.t   $\n(a,r_a+1)$.
Thus,  $p_{i,ab}(z)=\sum_{j=1}^{n_b}z^{j-1}A_{\bar\nu,\bar\mu,\n(b,j)}^{(II,\n(a,r_a+1))}(z^{n_b})$.

\end{enumerate}
  \end{pro}

The reader should notice that evolution of the weights given trough \eqref{evol.mg} give the following evolution of $w_{1,jb}$ and $w_{2,kc}$
\begin{align*}
  w_{1,jb}(t)&=\exp(\sum_{k\geq 1}y^{k/n_b}t_{kb})w_{1,jb},&
  w_{2,kc}(t)&=\sum_{l\geq 0} w_{2,k+l\; c}\bar s^{(c)}_l.
\end{align*}
Using \eqref{rho.evol} we may write
\begin{align*}
  w_{1,jb}(t)&=\exp(\sum_{k\geq 1}y^{k/n_b}t_{kb})w_{1,jb},\\
  w_{2,kc}(t)&=\exp(-\bar t_b^{(0)}(y^{1/m_c}))\sum_{l=0}^{m_b-1}\tilde w^{(l)}_{2k,c}\exp{(-\bar t_b^{[l]}(y^{1/m_c}))},&\tilde w^{(l)}_{2,kc}&:=\frac{1}{m_c}\sum_{i=1}^{m_b-1}\varepsilon_c^{il}y^{-i/m_c}w_{2\;k+i,c}.
\end{align*}
These evolved weights fulfill the positivity condition (recall that their support is included in the positive real line) when we have $t_{jb},\bar t_{jm_c\,c}\in\R$ and $\bar t_{jc}\leq 0$  for $j\neq 0$ $\text{mod} (m_c)$.

\subsection{ Riemann--Hilbert problems}
As we have just shown the generalized matrix polynomials are connected with a family of multiple orthogonal polynomials for a particular rank one moment matrix.
In \cite{daems-kuijlaars} the Riemann--Hilbert problem for multiple orthogonal polynomials of mixed type was presented, see also \cite{assche}. We will discuss its relation with the generalized orthogonal polynomials $p_i(z)$.

Let us recall the reader that the Cauchy transform is defined by
\begin{equation*}
    \hat Q_{\bar \nu' ,\bar \mu', K}(z):=-\frac{1}{2\pi
    i}\int_{\R}\frac{Q_{\bar \nu', \bar \mu'}(x)}{z-x}w_{2K}(x)\d x.
\end{equation*}

Now we can make the following definition for the $(p+q) \times (p+q)$
complex valued matrix $Y(z)$
\begin{align*}
Y_{K,J}&:=A_{\bar \nu'+e_K, \bar \mu', J}^{(II,K)} & J=1,\ldots,p \quad K=1,\ldots,p \\
Y_{K,J+p}&:=\hat Q_{\bar \nu'+e_K, \bar \mu, J}^{(II,K)} & J=1,\ldots,q \quad K=1,\ldots,p \\
Y_{K+p,J}&:=-2\pi i A_{\bar \nu', \bar \mu'-e_K, J}^{(I,K)} & J=1,\ldots,p \quad K=1,\ldots,q\\
Y_{K+p,J+p}&:=-2\pi i\hat Q_{\bar \nu', \bar \mu'-e_K, j}^{(I,K)} & J=1,\ldots,q \quad K=1,\ldots,q \\
\end{align*}
we will also use the following real valued  $(p+q) \times (p+q)$ matrix $D(x)$ defined by
 blocks
\begin{align*}
    D(x):=
    \begin{pmatrix}
    \I_p & W(x)\\
    0_{q \times p} & \I_q
    \end{pmatrix}
\end{align*}
where
$W_{JK}(x)=w_{1J}(x)w_{2K}(x)$.

We adapt to the present situation a result of \cite{daems-kuijlaars} taking into account the support of the weights

\begin{theorem}
Let be $\bar \nu', \bar \mu'$ two multi-indices such that $|\bar\nu'|=|\bar\mu'|$and
suppose that the normality conditions hold. Let also be two sets of
weight functions $\{w_{1J}\}_{J=1,\dots,p}$ and
$\{w_{2K}\}_{K=1,\dots,q}$ such that for every $J,K$
$w_{1J},(x)w_{2K}(x)$ are differentiable a.e. in $\R_+$ and $x^jw_{1J},x^jw_{2K}\in H^1(\R_+)$, $j=0,\dots,\nu_K'-1$.
In $x=0$ we require the weight functions to be bounded.
Then, the matrix $Y(z)$ is the only
solution of the following Riemann--Hilbert problem.
\begin{enumerate}
  \item $Y(z)$ is analytic in $\C \setminus \R_+$. \label{RH1}
  \item $Y(x)_+=Y(x)_-D(x)$ for all $x>0$. \label{RH2}
  \item $Y(z) \diag
  (z^{-\nu_1'},\ldots,z^{-\nu_p'},z^{\mu_1'},\ldots,z^{\mu_p'})=\I_{(p+q)}+O\left(z^{-1}\right)$
  for $z\rightarrow \infty$. \label{RH3}
  \item \label{RH4}$Y(z)=O \begin{pmatrix}1 & 1 & \cdots & 1 & \log |z| & \log |z| & \cdots & \log |z| \\
                               1 & 1 & \cdots & 1 & \log |z| & \log |z| & \cdots & \log |z| \\
                               \vdots & \vdots & \ddots  & \vdots & \vdots
                               & \vdots & \ddots  & \vdots \\
                               1 & 1 & \cdots & 1 & \log |z| & \log |z| & \cdots & \log |z|

                               \end{pmatrix}$ when $z \rightarrow 0$.
\end{enumerate}
\end{theorem}
\begin{proof}

 First we see uniqueness of the solutions. First we see that as $\det D(x)=1$ $\det Y(z)$ is analytical across the integration contour, then the only possible singularity of $\det Y(z)$ is in $z=0$. As $z \det Y(z) \rightarrow 0$ when $z \rightarrow 0$ the isolated singularity must be removable and $\det Y(z)$ is an entire function. The asymptotic behavior and Liouville's Theorem gives $ \det Y(z) =1 $. Consequently $Y(z)^{-1}$ exists. Given two possible solutions for the problem $Y(z)$ and $\tilde Y(z)$, the matrix $Y(z) \tilde Y(z)^{-1}$ can be singular only in $z=0$. As before the singularity must be removable so $Y(z) \tilde Y(z)^{-1}$ is entire and bounded, and hence $Y(z) \tilde Y(z)^{-1}=\I_{p+q}$.

 Now we prove that the matrix $Y(z)$ is a solution of the R--H problem. Condition \ref{RH1} follows from the fact that polynomials are entire functions and general theory of Cauchy integrals \cite{gakov} gives analytic behavior outside the integration contour. Now, observe that condition \ref{RH2} is a jumping condition on the positive real axis and is a consequence of Plemelj formulae. For condition \ref{RH3} we notice that $Y(z)_{ii}$ is a monic polynomial with degree $\nu'_i$ for $i=1,\dots,p$. We also see that the leading term of each $Y(z)_{ii}$ is $z^{-\mu'_i}$ for $i=p,\dots,(p+q)$ due to the orthogonality relations and the type I normalization. Consequently the diagonal elements of $Y(z) \diag
  (z^{-\nu_1'},\ldots,z^{-\nu_p'},z^{\mu_1'},\ldots,z^{\mu_p'})$ are equal to $1$ and the rest of them vanish like $\frac1z$ asymptotically. Finally, condition \ref{RH4} is a consequence of the behavior of the polynomials and the Cauchy transforms. The boundness condition of the weights at $z=0$ makes the Cauchy integrals to have at much $\log$-type singularities at $z=0$.
\end{proof}

The Theorem applies to our situation giving
\begin{pro}
If $\bar\nu'=\bar\nu-e_{\n(a,r_a+1)}$  and $\bar\mu'=\bar\mu$, with $\bar\nu$ and $\bar\mu$ as in Proposition \ref{identification}, we have $\Pi_{ij,ab}=Y_{\n(a,r_a+1),\n(b,j)}$ and
\begin{align*}
  p_{i,ab}(z)&=\sum_{j=1}^{n_b}z^{j-1}Y_{\n(a,r_a+1),\n(b,k)}(z^{n_b}),\\
  Y_{\n(a,r_a+1),\n(b,j)}(z^{n_b})&=\frac{1}{n_b z^{j-1}}\sum_{l=0}^{n_b-1} \epsilon_b^{-l(j-1)}p_{i,ab}(\epsilon_b^{j}z).
\end{align*}
\end{pro}

We observe that in \cite{adler-vanmoerbeke} and \cite{adler-vanmoerbeke3} these generalized polynomials are considered for the one component case $N=1$ and for $n=m$ and therefore fit as a particular example of our polynomials and consequently are related with multiple orthogonal polynomials of mixed type. Apparently they  might  noticed this fact   later, as in \cite{adler-vanmoerbeke2} they claim that they considered multiple orthogonal polynomials in \cite{adler-vanmoerbeke}. However, we must say that the Riemann--Hilbert problem derived there is different from the Daems--Kuijlaars problem  considered here for multiple orthogonal polynomials. In fact,  the matrix $Y$  considered  in \cite{adler-vanmoerbeke} is not analytic in the upper-half plane and hence fails to satisfy the Riemann--Hilbert problem posed there.

\section*{Acknowledgements}
MM thanks economical support from the Spanish Ministerio de Ciencia e Innovaci\'{o}n, research project  FIS2008-00200 and UF
thanks economical support from the Spanish Ministerio de Ciencia e Innovaci\'{o}n research projects MTM2006-13000-C03-02 and MTM2007-62945 and from Comunidad de Madrid/Universidad Carlos III de Madrid project CCG07-UC3M/ESP-3339. MM reckons different and clarifying discussions with Dr. Mattia Caffasso.

\end{document}